\documentclass[usenatbib]{mnras}
\usepackage[T1]{fontenc}
\usepackage{ae,aecompl}
\usepackage{bm}
\usepackage{amsmath,amssymb}
\usepackage{dcolumn}
\usepackage[dvips]{graphicx}
\usepackage{epsf}
\usepackage{color}
\usepackage{epsf}
\usepackage{epsfig}
\usepackage{graphicx}
\usepackage{longtable}
\usepackage{psfrag}
\usepackage{txfonts}
\usepackage[Symbol]{upgreek}       %package that allows to write the "roman pi"
\usepackage{ulem}
\usepackage[usenames,dvipsnames]{xcolor}

\newcommand{\WISC}{WI$\times$SC}

\title{The dipole anisotropy of WISE~$\times$~SuperCOSMOS number counts}

\author[C. A. P. Bengaly Jr. et al.]{
C. A. P. Bengaly Jr.,$^{1,2}$\thanks{E-mail: carlosap87@gmail.com}
C. P. Novaes$^{2}$, %\thanks{E-mail: camilanovaes@on.br}
H. S. Xavier$^{3}$, %\thanks{E-mail: hsxavier@if.usp.br}
M. Bilicki$^{4,5,6}$, %\thanks{Email: bilicki@strw.leidenuniv.nl}
A. Bernui$^{2}$, %\thanks{E-mail: bernui@on.br}
\and
J. S. Alcaniz,$^{2,7}$%\thanks{E-mail: alcaniz@on.br}
\vspace{0.4cm}\\
$^{1}$Department of Physics and Astronomy, University of Western Cape, 7535, Cape Town, South Africa \\              
$^{2}$Coordena\c c\~ao de Astronomia e Astrof\'isica, Observat\'orio Nacional, 20921-400, Rio de Janeiro - RJ, Brazil \\ 
$^{3}$Instituto de Astrof\'isica, Geof\'isica e Ci\^encias Atmosf\'ericas, Universidade de S\~ao Paulo, 05508-090, S\~ao Paulo - SP, Brazil \\
$^{4}$Leiden Observatory, Leiden University, 2333, P.O. Box 9513, NL-2300 RA Leiden, The Netherlands \\          
$^{5}$National Center for Nuclear Research, Astrophysics Division, P.O. Box 447, 90-950, \L{}\'{o}d\'{z}, Poland \\
$^{6}$Janusz Gil Institute of Astronomy, University of Zielona G\'{o}ra, ul. Szafrana 2, 65-516 Zielona G\'ora, Poland \\
$^{7}$Physics Department, McGill University, Montreal QC, H3A 2T8, Canada\\
}

\begin{document}
\label{firstpage}
\pagerange{\pageref{firstpage}--\pageref{lastpage}}
\maketitle

%%%%%%%%%%%%%%%%%%%%%%%%%%%%%%%%%%%%%%%%%%%%%%%%%%%%%%%%%%%%%%%%%%%%%%%%%%%%%%%%%%%%%%%%%%%%%%%%%%%%%%%%%%%%%%%%%%%%%%%%%%%%%%%%%%%%%%%%%

\begin{abstract}
We probe the isotropy of the Universe with the largest all-sky photometric redshift dataset currently available, namely WISE~$\times$~SuperCOSMOS. We search for dipole anisotropy of galaxy number counts in multiple redshift shells within the $0.10 < z < 0.35$ range, for two subsamples drawn from the same parent catalogue. Our results show that the dipole directions are in good agreement with most of the previous analyses in the literature, and in most redshift bins the dipole amplitudes are well consistent with $\Lambda$CDM-based mocks in the cleanest sample of this catalogue. In the $z<0.15$ range, however, we obtain a persistently large anisotropy in both subsamples of our dataset. Overall, we report no significant evidence against the isotropy assumption in this catalogue except for the lowest redshift ranges. The origin of the latter discrepancy is unclear, and improved data may be needed to explain it.
\end{abstract}


%%%%%%%%%%%%%%%%%%%%%%%%%%%%%%%%%%%%%%%%%%%%%%%%%%%%%%%%%%%%%%%%%%%%%%%%%%%%%%%%%%%%%%%%%%%%%%%%%%%%%%%%%%%%%%%%%%%%%%%%%%%%%%%%%%%%%%%%%

\section{Introduction}
\label{sec:intro}

The current standard model of cosmology, called $\Lambda$CDM, assumes Friedmann-Lema\^{i}tre-Robertson-Walker as its background metric, and that the Universe is approximately homogeneous and isotropic on large scales, a feature of the so-called `Cosmological Principle' (CP). Despite the good agreement between $\Lambda$CDM and a plethora of cosmological observations~\citep[e.g.][]{planck16, sdss16}, direct tests of the CP need to be performed in order to assess whether it is a valid cosmological assumption or just mathematical simplification. Persistent lack of isotropy or homogeneity on large scales would require a complete reformulation of the current cosmological scenario, and thus of our understanding of the Universe.

It is well accepted that the spatial distribution of cosmic objects becomes statistically homogeneous on scales around $100-150$ Mpc/h~\citep{hogg05, scrimgeour12, pandey16, laurent16, ntelis17, goncalves17}. The only major dipole anisotropy, observed in the cosmic microwave background, is in the standard framework interpreted as an imprint of our own peculiar motion, rather than actual cosmological signal~\citep{kogut93, planck14}. We will refer to it as the kinematic dipole hereafter~\citep{maartens17}. However, probing this quantity is not our goal here, and we look instead for a number count dipole in the intrinsic galaxy distribution\footnote{For simplicity, we look for the largest-order anisotropic mode, which is a dipole in the number counts of galaxies.}. The observed galaxy count dipole is not expected to be aligned with the kinematic one since the intrinsic galaxy density fluctuations dominate over this signal on the scales of $z < 1$~(\citealt{gibelyou12, yoon15}). However, the presence of a larger dipole than predicted by these intrinsic fluctuations in the standard model would thus indicate deviations from isotropy or homogeneity. 

Therefore, all-sky infrared and optical catalogues are ideal probes for this test. This was previously performed by~\cite{itoh10, gibelyou12, appleby14, yoon14, alonso15, bengaly17}, and none of these papers reported compelling signal against the CP. In this work, we reassess the isotropy of the Universe on the $z<0.5$ scales using the WISE~$\times$~SuperCOSMOS catalogue~\citep[][hereafter \WISC]{bilicki16}\footnote{\url{http://ssa.roe.ac.uk/WISExSCOS}}, which is currently the largest and deepest all-sky photometric redshift (photo-$z$) dataset available. We check for concordance between the number count dipole in \WISC\ and in synthetic datasets assuming real data specifications, in addition to the power spectrum of the $\Lambda$CDM model. We therefore extend the analysis of~\cite{bengaly17} where another WISE-based catalogue was used, namely WISE-2MASS (W2M,~\citealt{kovacs15}), which not only was shallower than \WISC, but also did not include redshift information, and it comprised 10 times fewer sources. If found, strong discrepancies between the observational data and their respective mocks would hint at potential evidence against the cosmic isotropy assumption, unless we are restricted by persisting systematics.

%%%%%%%%%%%%%%%%%%%%%%%%%%%%%%%%%%%%%%%%%%%%%%%%%%%%%%%%%%%%%%%%%%%%%%%%%%%%%%%%%%%%%%%%%%%%%%%%%%%%%%%%%%%%%%%%%%%%%%%%%%%%%%%%%%%%%%%%%

\vspace{-0.5cm}

\section{Data selection}
\label{sec:data_selection}

\begin{figure}%[!t]
\includegraphics[scale=0.36]{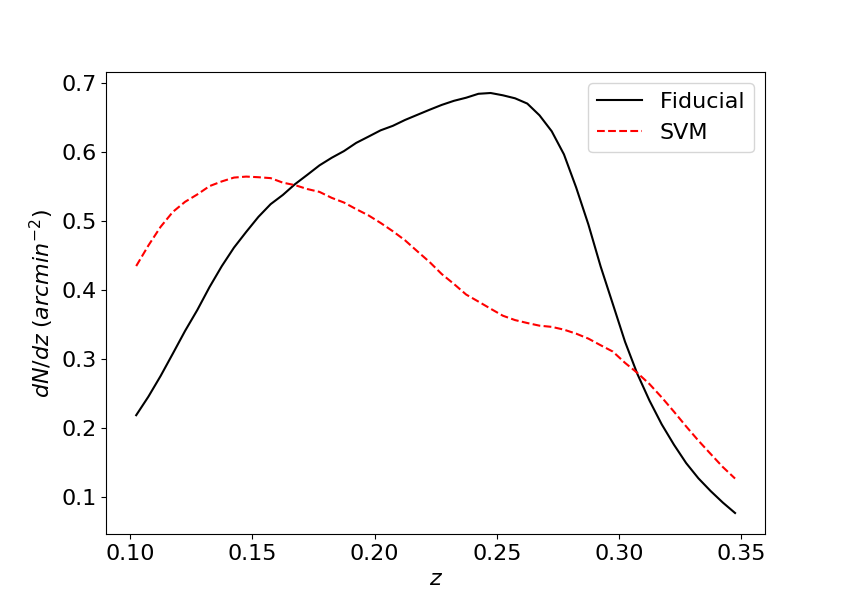}
\caption{The redshift distribution of the \WISC\ Fiducial (red dashed curve) and of the SVM sample (black solid curve), both given in counts per square arcminute per redshift bin.}
\label{fig:dn_dz}
\end{figure}

\begin{figure*}%[!t]
\includegraphics[width=0.45\linewidth, height=4.5cm]{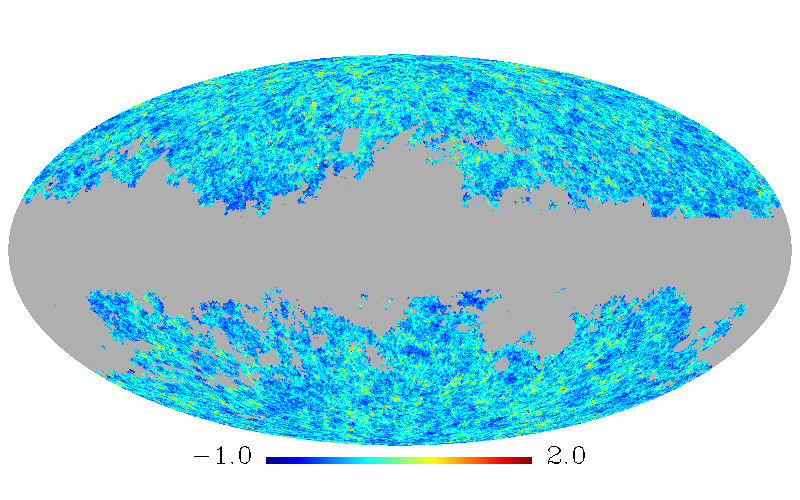}
\includegraphics[width=0.45\linewidth, height=4.5cm]{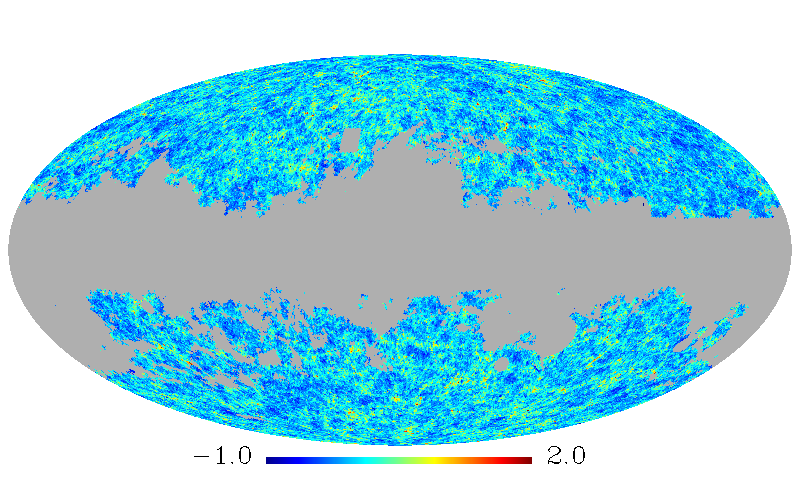}
\caption{{\it Left panel:} The density contrast of galaxy number counts (clipped at $\delta_\mathrm{max}=2.0$ to ease visualisation) of the \WISC\ Fiducial sample in the $0.10 < z < 0.35$ range, i.e., the full sample analysed here. {\it Right panel:} Same as the left panel, but for the SVM sample. The grey area corresponds to the masked region as discussed in section~\ref{sec:data_selection}.}
\label{fig:numbercountmap_cumul}
\end{figure*}

\begin{figure*}%[!ht]
\includegraphics[width=0.45\linewidth, height=4.5cm]{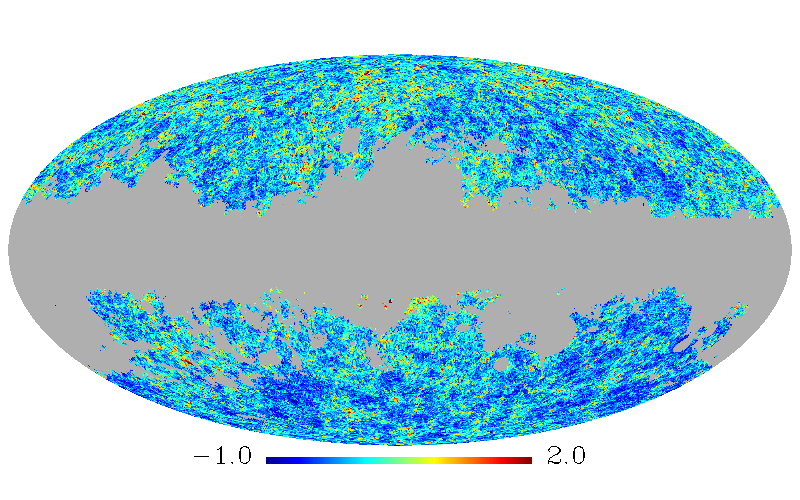}
\includegraphics[width=0.45\linewidth, height=4.5cm]{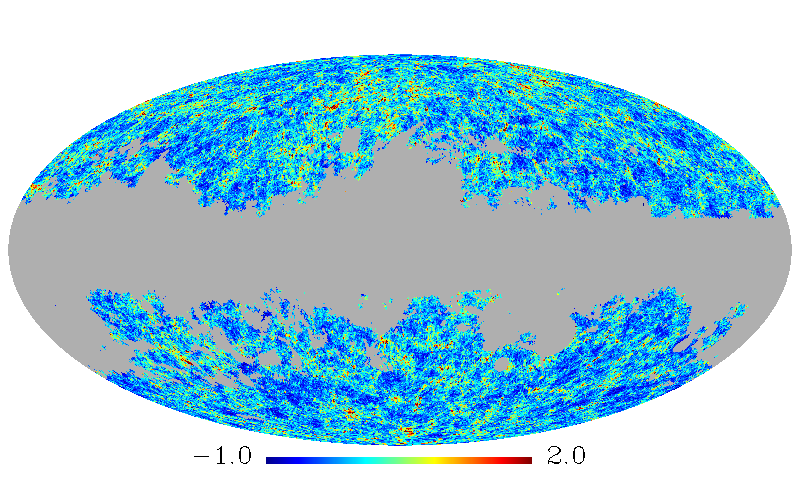}
\caption{Same as Fig.~\ref{fig:numbercountmap_cumul}, but for galaxies within $0.10 < z \leq 0.15$ only.}
\label{fig:numbercountmap_tomog}
\end{figure*}

The \WISC\ photo-$z$ catalogue~\citep{bilicki16} is based on a cross-match of two photometric all-sky samples, WISE~\citep{wright10} and SuperCOSMOS~\citep{peacock16}. This dataset is flux-limited to $B<21$, $R<19.5$, and $13.8 < W1 < 17$ (3.4 $\mu$m, Vega) and provides photo-$z$s for all the included sources, ranging from $0<z<0.4$ (mean $\langle z \rangle \simeq 0.2$) with typical photo-$z$ error $\sigma_{\delta z} = 0.033(1+z)$. The data come with a fiducial mask which removes low Galactic latitudes ($|b| \leq 10^{\circ}$ up to $|b| \leq 17^{\circ}$ by the Bulge), areas of high Galactic extinction ($E(B-V) > 0.25$), as well as other contaminated regions. Here we however apply more strict cuts to avoid selection effects due to extinction, namely $E(B-V) > 0.10$, and require $0.10 < z_\mathrm{phot} < 0.35$ to remove low-redshift prominent structures as well as the high-redshift tail of \WISC, where the data is very sparse.

The original \WISC\ data have had specific galaxy selection applied to remove stellar and quasar contamination (the latter being minimal). As purity is more important for our purposes than completeness, we applied a more aggressive colour cut than originally in \cite{bilicki16}, namely  $W1 - W2 > 0.2$ over the entire sky, which should guarantee very efficient star removal~\citep{jarrett17}. We will call this \WISC\ sample with the additional cleanup `Fiducial' from now on. In an alternative approach to galaxy identification in \WISC,~\cite{krakowski16} used the Support Vector Machines classification algorithm which separates sources in multi-colour space using best-fit hyperplanes, and obtained a purer galaxy sample than the main \WISC\ one, yet less complete. This `SVM' dataset comes with probability estimates of sources belonging to a given class, and we applied a conservative lower cutoff of $p_{gal}>2/3$, so that our selected objects have at least 67\% of probability of being a galaxy rather than any other class (star or AGN). This cut results in $p_{gal, mean} \simeq 0.90$ in each redshift bin, besides $>$35\% of the SVM sources with $p_{gal}>0.95$. 

After applying the \WISC\ mask as well as our additional cuts on $E(B-V)$ and photo-$z$'s, we obtained samples of 9.5 and 8.3 million galaxies over $f_{sky} \simeq 0.545$ for the Fiducial and SVM datasets, of median redshifts $z\mathrm{^{Fid}_{med}} \simeq 0.22$ and $z\mathrm{^{SVM}_{med}} \simeq 0.20$. Depending on the redshift range, these two \WISC\ subsets typically have $\sim 50$--$75\%$ common sources. We show the redshift distribution of both datasets in Fig.~\ref{fig:dn_dz}; number count maps of these samples for the full redshift range ($0.10 < z < 0.35$) are featured in Fig.~\ref{fig:numbercountmap_cumul}, and Fig.~\ref{fig:numbercountmap_tomog} exhibits objects in the $0.15 < z \leq 0.20$ bin only. All maps were produced using HEALPix~\citep{gorski05} with resolution of $N_{side}=128$ (pixel size of $\sim 0.5^{\circ}$). 

\vspace{-0.5cm}

%%%%%%%%%%%%%%%%%%%%%%%%%%%%%%%%%%%%%%%%%%%%%%%%%%%%%%%%%%%%%%%%%%%%%%%%%%%%%%%%%%%%%%%%%%%%%%%%%%%%%%%%%%%%%%%%%%%%%%%%%%%%%%%%%%%%%%%%%

\section{Methodology}
\label{sec:methodology}

The isotropy of galaxy number counts is estimated with the delta-map method first presented in~\cite{alonso15} (see also~\citealt{bengaly17}), in which the sky is decomposed into 768 large HEALPix pixels ($N_{\rm side}=8$), and hemispheres are constructed using the respective pixel centres as symmetry axes. The delta-map is then computed as

\begin{equation}
\label{eq:delta_n}
\Delta_i = 2 \times \left( \frac{n_i^U - n_i^D}{n_i^U + n_i^D} \right)\;,
\end{equation}

\noindent where $n_i^j \equiv N^j_i/(4\pi f^j_{sky,i})$ are counts in the $i$-th hemisphere, $i\in{1, ..., 768}$, $j$ represents the hemispheres indexes ``up'' ($U$) and ``down'' ($D$) defined according to this pixellisation scheme, whereas $N^j_i$ and $f^j_{sky,i}$ are the total number of objects and the observed fraction of the sky encompassed in each of these hemispheres, respectively.

The dipole of galaxy number counts is obtained by expanding the delta-map from~\eqref{eq:delta_n} into spherical harmonics. From the $\{a_{\ell m}\}$, coefficients we select the $\ell = 1$ terms, i.e., the $\{a_{1m}\}$, to reconstruct only the dipole component of the delta-map, $\Delta_{dip} = \sum a_{1m} Y_{1m}$. Therefore, we quote the maximum value of the $\Delta_{dip}$ map as our dipole amplitude $A$, in addition to the direction where it points to. In this work, the \WISC\ catalogue is additionally decomposed into redshift shells before the delta-map calculation: cumulative ones, i.e., $0.10 < z \leq 0.15; 0.10 < z \leq 0.20; ...; 0.10 < z < 0.35$, and disjoint ones, $0.10 < z \leq 0.15; 0.15 < z \leq 0.20; ...;$ etc. 

The statistical significance of the delta-map dipoles is calculated from \WISC\ mock catalogues produced with the {\sc flask} code\footnote{\url{http://www.astro.iag.usp.br/~flask}}~\citep{xavier16}. These mocks are full-sky lognormal realisations of the density field in redshift shells based on the input angular power spectra $C_{\ell}^{(z_i z_j)}$ ($z_i$ and $z_j$ denoting different redshift shells) provided by {\sc camb sources}~\citep{challinor11}, which are Poisson-sampled according to the \WISC\ selection function. The input $C_{\ell}^{(z_i z_j)}$ were computed for redshift distributions that are convolutions of the $\Delta z = 0.05$ shells with Gaussian scatter of $\sigma_{\delta z} = 0.033(1+z)$ (representing \WISC\ photo-z errors) using $\Lambda$CDM best-fit parameters~\citep{planck16}, and they include linear redshift space distortions, gravitational lensing distortions of the volume elements, and non-linear contributions modelled by {\sc halofit}~\citep{smith03, takahashi12}. We applied a linear scaling factor to each $C_{\ell}^{(z_i z_j)}$ -- playing a role similar to galaxy bias -- which was used to match the variances of counts in pixels to the ones observed in the real data.

We additionally compared the Fiducial dataset source distribution to SDSS~\citep{york00} in a $\mathrm{1^\circ}$-wide strip centred on declination $\delta=\mathrm{30^\circ}$ and estimated that it still contained a fraction $f_{\mathrm{star}}$ of stars that is well fitted by $f_{\mathrm{star}}=0.71\exp(-0.09|b|)+0.013$. Therefore, we Poisson-sampled stars according to this distribution and included them in our mocks. By adjusting the selection function normalisation and the $C_{\ell}^{(z_i z_j)}$ scaling factors, we made our simulations match the Fiducial dataset in terms of $f_{\mathrm{star}}$, mean number of objects (galaxies + stars) and variance in the pixels\footnote{The simulation input files are available at:\\ \url{http://www.astro.iag.usp.br/~flask/sims/wisc17.tar.gz}}.

Following the prescriptions above, we produced 1000 full-sky mocks of both Fiducial and SVM datasets in each $\Delta z=0.05$ photo-z bin, spanning the $0.10 < z < 0.35$ range, using the same resolution as for the real data maps ($N_{\rm side}=128$). From these realisations, we computed how many of them featured a dipole amplitude at least as large as the real data for each $z$-bin analysed, hereafter quoted as $p$-values. Low $p$-values, such as $p<0.005$~\citep{planck16}, will be regarded as an indication that the model cannot fully describe the observations, and thus might be interpreted as challenging the concordance model. 

%%%%%%%%%%%%%%%%%%%%%%%%%%%%%%%%%%%%%%%%%%%%%%%%%%%%%%%%%%%%%%%%%%%%%%%%%%%%%%%%%%%%%%%%%%%%%%%%%%%%%%%%%%%%%%%%%%%%%%%%%%%%%%%%%%%%%%%%%

\vspace{-0.5cm}

\section{Results}
\label{sec:results}

\begin{figure*}%[!t]
\includegraphics[width=0.45\linewidth, height=4.3cm]{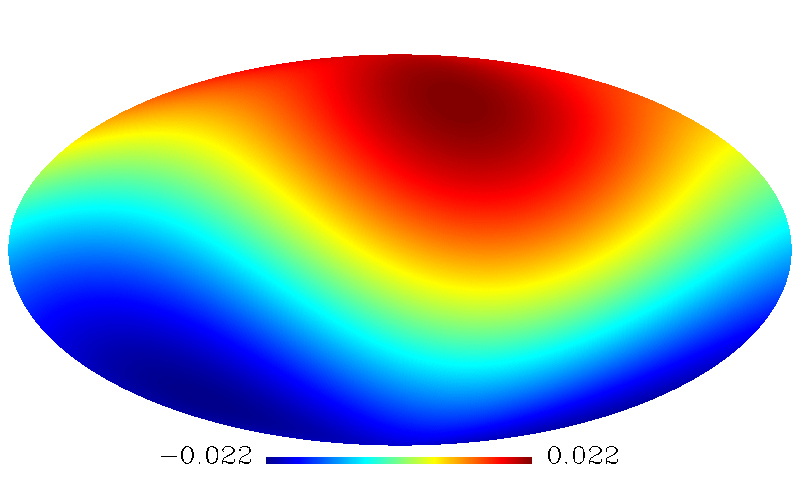}
\includegraphics[width=0.45\linewidth, height=4.3cm]{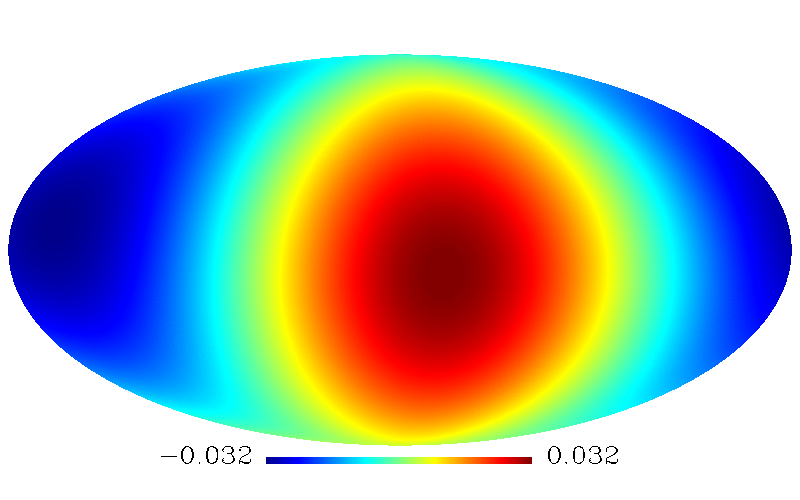}
\caption{{\it Left panel:} The Delta-map dipole of the \WISC\ Fiducial data set in the $0.10 < z < 0.35$ range. {\it Right panel:} Same as the left panel, but for the \WISC\ SVM galaxies. Both maps are represented in Galactic coordinates. Further details on the amplitudes and directions, as well as on the other redshift bins, are presented in Tab.~\ref{tab:dipoles}.}
\label{fig:deltamap_cumul}
\end{figure*}

\begin{table}%[!t]
%\begin{center}
\begin{tabular}{ccccc} 
\hline
\hline
$z$-bin (Fiducial) & $A \times 10^{-1}$ & $(l,b)$ & $\sigma(\theta)$ & $p$-value \\
\hline
$\mathbf{0.10 < z \leq 0.15}$ & $\mathbf{1.474(18)}$ & $\mathbf{(56.5^{\circ},69.8^{\circ})}$ & $\mathbf{0.6^{\circ}}$ & $\mathbf{<0.001}$ \\
$\mathbf{0.10 < z \leq 0.20}$ & $\mathbf{0.701(11)}$ & $\mathbf{(20.8^{\circ},70.2^{\circ})}$ & $\mathbf{0.7^{\circ}}$ & $\mathbf{<0.001}$ \\
$\mathbf{0.10 < z \leq 0.25}$ & $\mathbf{0.394(09)}$ & $\mathbf{(346.1^{\circ},69.8^{\circ})}$ & $1.0^{\circ}$ & $\mathbf{0.001}$ \\
$0.10 < z \leq 0.30$ & $0.250(07)$ & $(317.9^{\circ},61.2^{\circ})$ & $1.5^{\circ}$ & $0.084$  \\
$0.10 < z < 0.35$ & $0.225(07)$ & $(319.9^{\circ},59.7^{\circ})$ & $1.5^{\circ}$ & $0.129$ \\
\hline
$0.15 < z \leq 0.20$ & $0.303(13)$ & $(325.2^{\circ},43.8^{\circ})$ & $2.5^{\circ}$ & $0.129$ \\
$0.20 < z \leq 0.25$ & $0.200(10)$ & $(272.8^{\circ},-9.9^{\circ})$ & $6.9^{\circ}$ & $0.332$ \\
$0.25 < z \leq 0.30$ & $0.293(13)$ & $(262.6^{\circ},-42.2^{\circ})$ & $3.2^{\circ}$ & $0.017$ \\
$0.30 < z < 0.35$ & $0.112(22)$ & $(27.7^{\circ},-59.3^{\circ})$ & $9.7^{\circ}$ & $0.773$ \\
\hline
\hline
$z$-bin (SVM) & $A \times 10^{-1}$ & $(l,b)$ & $\sigma(\theta)$ & $p$-value \\
\hline
$\mathbf{0.10 < z \leq 0.15}$ & $\mathbf{0.863(14)}$ & $\mathbf{(29.3^{\circ},65.7^{\circ})}$ & $\mathbf{1.0^{\circ}}	$ & $\mathbf{<0.001}$ \\
$0.10 < z \leq 0.20$ & $0.417(08)$ & $(342.1^{\circ},26.3^{\circ})$ & $2.6^{\circ}$ & $0.019$ \\
$0.10 < z \leq 0.25$ & $0.371(06)$ & $(332.6^{\circ},-3.2^{\circ})$ & $2.8^{\circ}$ & $0.010$ \\
$0.10 < z \leq 0.30$ & $0.320(06)$ & $(335.0^{\circ},-7.1^{\circ})$ & $2.7^{\circ}$ & $0.010$ \\
$0.10 < z < 0.35$ & $0.316(06)$ & $(339.6^{\circ},-9.7^{\circ})$ & $2.7^{\circ}$ & $0.007$ \\
\hline
$\mathbf{0.15 < z \leq 0.20}$ & $\mathbf{0.674(14)}$ & $\mathbf{(315.4^{\circ},-34.2^{\circ})}$ & $\mathbf{1.5^{\circ}}$ & $\mathbf{<0.001}$ \\
$\mathbf{0.20 < z \leq 0.25}$ & $\mathbf{0.682(16)}$ & $\mathbf{(310.5^{\circ},-52.4^{\circ})}$ & $\mathbf{1.2^{\circ}}$ & $\mathbf{<0.001}$ \\
$0.25 < z \leq 0.30$ & $0.166(17)$ & $(13.3^{\circ},-48.5^{\circ})$ & $5.1^{\circ}$ & $0.236$ \\
$\mathbf{0.30 < z < 0.35}$ & $\mathbf{0.370(19)}$ & $\mathbf{(18.6^{\circ},-18.8^{\circ})}$ & $\mathbf{8.7^{\circ}}$ & $\mathbf{<0.001}$ \\
\hline
\hline
\end{tabular}
\caption{The amplitude (and its respective error bar down to two significant digits; col.2), direction (col.3), uncertainty on the direction (col.4), and statistical significance (given in $p$-values; col.5) of the \WISC\ dipole obtained from the Fiducial (top) and SVM (bottom) samples. We highlight the cases where significant deviation from isotropy according to the $p<0.005$ criterion was found.}
\label{tab:dipoles}
%\end{center}
\end{table}

\begin{table*}
\resizebox{\textwidth}{!}{%
\begin{tabular}{cccccccccc}
\hline
\hline
Dipole &  \WISC\ Fid full & \WISC\ SVM full & W2M-17 & 2MPZ-15 & W2M-14 & 2MPZ-14 & \WISC\ Fid low-z & \WISC\ SVM low-z \\
\hline
\hline
\WISC\ Fid full    & - & $0.227$ & $0.247$ & $0.249$ & $0.275$ & $0.263$ & $0.815$ & $0.764$ \\
\WISC\ SVM full    & $0.227$ & - & $0.007$ & $0.013$ & $0.013$ & $0.091$ & $0.253$ & $0.239$ \\
W2M-17             & $0.247$ & $0.007$ & - & $0.005$ & $0.016$ & $0.072$ & $0.287$ & $0.259$ \\
2MPZ-15            & $0.249$ & $0.013$ & $0.005$ & - & $0.022$ & $0.067$ & $0.377$ & $0.340$ \\
W2M-14             & $0.275$ & $0.013$ & $0.016$ & $0.022$ & - & $0.076$ & $0.217$ & $0.186$ \\
2MPZ-14            & $0.263$ & $0.091$ & $0.072$ & $0.067$ & $0.076$ & - & $0.580$ & $0.541$ \\   
\WISC\ Fid low-z   & $0.815$ & $0.253$ & $0.287$ & $0.377$ & $0.217$ & $0.580$ & - & $0.233$ \\
\WISC\ SVM low-z   & $0.764$ & $0.239$ & $0.259$ & $0.340$ & $0.186$ & $0.541$ & $0.233$ & - \\
\hline
\hline
\end{tabular}}
\caption{The probability that randomly picked directions are closer to each other than the measured distance between the corresponding dipoles, where small values indicate strongly aligned directions. \WISC\ Fid full and \WISC\ SVM full correspond to the Fiducial and SVM samples analysed in the $0.10 < z < 0.35$ range, respectively, while both \WISC\ Lowz cases represent the $0.10 < z \leq 0.15$ bins. The remaining directions were obtained using the WISE-2MASS and 2MASS photo-z datasets (respectively W2M and 2MPZ), as reported by~\protect\cite{bengaly17, alonso15, yoon14, appleby14} for W2M-17, 2MPZ-15, W2M-14, and 2MPZ-14, respectively.}
\label{tab:angdist_prob}
\end{table*}

The dipoles resulting from the delta-map analyses of the two \WISC\ samples are shown for the full redshift range in Fig.~\ref{fig:deltamap_cumul}, while the dipole directions and amplitudes for each redshift bin are presented in Table~\ref{tab:dipoles} for both Fiducial and SVM datasets. The error bars of $A$ and $(l,b)$ were estimated from 1000 simulations with fixed underlying density fluctuations (and therefore fixed dipole direction and amplitude), but with different Poisson noise, for each $z$-bin. Then, $\sigma_A$ corresponds to the average deviation from the fiducial dipole value, i.e., the quoted $A$, and $\sigma(\theta)$ denotes the average angular distance from the observed dipole direction, as given by this set of simulations. We readily verify that the dipole amplitude decreases when we probe the number counts at larger depths, as it goes from $A \simeq 0.10$ in the thinnest ($0.10 < z \leq 0.15$) to $A \simeq 0.03$ in the thickest ($0.10 < z < 0.35$) cumulative shell, while the deepest shell ($0.30 < z < 0.35$) gives $A = 0.01-0.04$ depending on the sample. This behaviour was also noted by~\cite{yoon14} when comparing their results with those in~\cite{gibelyou12}, which used shallower catalogues in some of their analyses. This amplitude decrease with increased redshift is due to the smoother rms density fluctuations in the large-scale structure in higher $z$, in addition to the increasing volume probed in these larger cumulative shells. 

The comparison between the dipole amplitude of the actual observations and lognormal \WISC\ mocks, for the cumulative bins, shows only marginal agreement for the SVM sample in the two thickest bins, while the Fiducial sample performs slightly better, reaching $p = 0.129$ in the largest redshift shell. When analysing the tomographic bins, on the other hand, we find good consistency between the Fiducial data and its mocks, however the performance of the SVM sample was good only in the $0.25 < z \leq 0.30$ shell. Moreover, we find that both SVM and Fiducial data show a larger dipole amplitude than the mocks in the lowest redshift shell, $0.10 < z \leq 0.15$, which interestingly is the bin presenting the best mutual agreement between these two samples. 

However, we also note that the concordance between data and simulations improves when the cumulative redshift shells encompass more distant galaxies. This is more evident for the Fiducial sample, where $p > 0.05$ in the two thickest redshift shells, yet less than $\sim 2\%$ of the SVM realisations have $A$ larger than the real data in the same ranges. In the tomographic $z$-bins, we found that the Fiducial data is in good concordance with its $\Lambda$CDM-based mocks except for the $0.25 < z \leq 0.30$ bin which, interestingly, is the redshift shell where the SVM dipole amplitude agrees best with simulations. From these results, we conclude that the Fiducial dataset shows better concordance with its respective mocks than the SVM one, and that the \WISC\ data agree with the isotropy hypothesis of the Universe apart from the lowest redshift ranges examined here. At present, we cannot unambiguously resolve the reason for the latter discrepancy, and better all-sky data covering these redshifts will be needed for that purpose.

We also stress that the directions of the number count dipoles in cumulative redshift shells, especially in the range $0.10 < z < 0.35$ of the SVM dataset, are in good agreement with similar analyses in the literature. We calculate the probabilities $P_\theta$ that the alignments between the corresponding directions would occur at random such as

\begin{equation}
P_\theta = \frac{1}{4\pi}\int_0^\theta2\pi\sin\theta'\mathrm{d}\theta' = \frac{1-\cos\theta}{2},
\end{equation}

\noindent where $\cos{\theta} = \sin{b_1}\sin{b_2} + \cos{b_1}\cos{b_2}\cos{(l_1 - l_2)}$. A quantitative assessment of the concordance between these dipole directions is provided in Tab.~\ref{tab:angdist_prob}. For instance,~\cite{bengaly17} obtained a dipole anisotropy with $A = 0.0507$ pointing towards $(l,b) = (323^{\circ},-5^{\circ})$ in the W2M catalogue which peaks at $z \sim 0.14$, which is consistent with the results from~\cite{yoon14}, i.e. $(l,b) = (310^{\circ}, -15^{\circ})$. Furthermore, \cite{appleby14} and \cite{alonso15} found, using different methods, preferred directions of respectively $(l,b) = (320^{\circ},6^{\circ})$ and $(l,b) = (315^{\circ},30^{\circ})$ in the 2MASS Photo-$z$ dataset (2MPZ,~\citealt{bilicki14})

%%%%%%%%%%%%%%%%%%%%%%%%%%%%%%%%%%%%%%%%%%%%%%%%%%%%%%%%%%%%%%%%%%%%%%%%%%%%%%%%%%%%%%%%%%%%%%%%%%%%%%%%%%%%%%%%%%%%%%%%%%%%%%%%%%%%%%%%%

\vspace{-0.6cm}

\section{Conclusions}
\label{sec:conclusions}

In this work, we probed the cosmological isotropy at the $0.10 < z < 0.35$ interval through the directional dependence of the galaxy counts in the WISE~$\times$~SuperCOSMOS catalogue. To do so, we adopted a hemispherical comparison method, and its dipole contribution provided our diagnostic of cosmological anisotropy. The observational samples consisted of two datasets, namely `Fiducial' and `SVM', which differ in how galaxies were identified in them: through colour cuts in the former, and by means of automatised classification in the latter. Thanks to the availability of redshift information, we were able to perform this test in tomographic $z$-bins, which enabled a natural extension of the analysis carried out in~\cite{bengaly17} with the shallower WISE-2MASS sample. 

We found overall good agreement between the \WISC\ dipole directions obtained here and those from previous analyses in the literature. As far as the dipole amplitudes are concerned, their level of agreement with $\Lambda$CDM mocks is different for various redshift shells and sample selections. In both Fiducial and SVM cases, the lowest-redshift bin $0.10<z<0.15$ is discrepant from the mocks; at higher redshifts, the Fiducial sample exhibits good agreement with the simulations, which is generally not the case for the SVM one. Interestingly, the $z<0.15$ range is the only case in which both samples agree with each other regarding the dipole direction. In all other cases, the dipole amplitudes and directions significantly differ between Fiducial and SVM selections in individual $z$-shells. Although there is some interception of roughly $\sim 50-75$\% between the two \WISC-based datasets, they rely upon different methods to separate galaxies from stars and quasars, resulting in distinct samples from the same catalogue, and perhaps selecting different galaxy types for each one of them. However, it is very unlikely that this difference would explain such effect, as no significant colour discrepancy was found between the two samples. 

Even if we credit the better agreement between the Fiducial sample and the lognormal realisations to a more rigorous criterion to eliminate stars, as described in Sec.~\ref{sec:data_selection}, or to different galaxy types, this procedure still cannot explain the large $A$ obtained for both preselections in the $0.10 < z \leq 0.15$ redshift shell. This result could be an indication of either related systematics in both datasets, or the presence of very large, local density fluctuations, which can increase the number counts dipole as pointed out by~\cite{rubart14}. A more thorough investigation of these hypotheses will be pursued in the future, but it may require better all-sky datasets, which at present are not available for the relevant redshift ranges. 

This work presents the first contribution of the \WISC\ catalogue to Cosmology in the form of an updated test of the large-scale isotropy of the Universe, in which we found no significant departure from this fundamental hypothesis for $z > 0.15$, yet we are still very limited by the completeness and systematics of the available data. Nonetheless, the \WISC\ data set can be considered a testbed for forthcoming surveys, especially LSST~\citep{lsst09} and SKA~\citep{ska15}, as they will reach much deeper scales on large sky areas and, therefore, will enable much more precise tests of the CP in the years to come~\citep{itoh10, yoon15}.

%%%%%%%%%%%%%%%%%%%%%%%%%%%%%%%%%%%%%%%%%%%%%%%%%%%%%%%%%%%%%%%%%%%%%%%%%%%%%%%%

\vspace{-0.5cm}

\section*{Acknowledgements}

CAPB acknowledges South African SKA Project, besides CAPES for financial support in the early stage of this work. CPN is supported by the DTI-PCI Programme of the Brazilian Ministry of Science, Technology, Innovation and Communications (MCTIC). HSX acknowledges FAPESP for financial support. MB is supported by the Netherlands Organization for Scientific Research, NWO, through grant number 614.001.451, and by the Polish National Science Center under contract \#UMO-2012/07/D/ST9/02785. AB thanks the Capes PVE project 88881.064966/2014-01. JSA is supported by CNPq and FAPERJ. We thank the Wide Field Astronomy Unit at the Institute for Astronomy, Edinburgh, for archiving the WISE~$\times$~SuperCOSMOS catalogue. We also acknowledge using the {\sc HEALPix} package for the derivation of the results presented in this work. 

%%%%%%%%%%%%%%%%%%%%%%%%%%%%%%%%%%%%%%%%%%%%%%%%%%%%%%%%%%%%%%%%%%%%%%%%%%%%%%%%

%%%%%%%%%%%%%%%%%%%%%%%%%%%%%%%%%%%%%%%%%%%%%%%%%%%%%%%%%%%%%%%%%%%%%%%%%%%%%%%%%%%%%%%%%%%%%%%%%%%%%%%%%%%%%%%%%%%%%%%%%%%%%%%%%%%%%%%%%

% Don't change these lines
\bsp	% typesetting comment
\label{lastpage}

\end{document}